\documentclass[journal]{IEEEtran}
\IEEEoverridecommandlockouts
\pdfoutput=1
\usepackage{cite}
\usepackage{amsmath,amssymb,amsfonts}
\usepackage{algorithm,algorithmic}
\usepackage{bm,bbm}
\usepackage{booktabs}
\usepackage{graphicx}
\usepackage{textcomp}
\usepackage{xcolor}
\usepackage{subfigure}

\newtheorem{pb}{Problem}

\def\BibTeX{{\rm B\kern-.05em{\sc i\kern-.025em b}\kern-.08em
		T\kern-.1667em\lower.7ex\hbox{E}\kern-.125emX}}
\begin{document}
	
	\title{Age Optimum Sampling in Non-Stationary Environment}

	\author{
		Jinheng~Zhang,~Haoyue~Tang,~Jintao~Wang, Sastry Kompella and~Leandros~Tassiulas\\
		
		\thanks{J. Zhang and J. Wang are with Beijing National Research Center for Information Science and Technology (BNRist) and the Department of Electronic Engineering, Tsinghua University, Beijing 100084, China. (e-mail: 	\{jinhang-20@mails.; wangjintao@\}tsinghua.edu.cn)}
		\thanks{H. Tang and L. Tassiulas are with the Department of Electrical Engineering and Yale Institute of Network Science, New Haven, CT, 06511, USA. (e-mail: \{haoyue.tang, leandros.tassiulas\}@yale.edu)}
		\thanks{S. Kompella was with the Naval Research Laboratory. He is now with Nexcepta Inc, Gaithersburg, MD 20878, USA. (e-mail: skompella@nexcepta.com) }
		
		\thanks{The work of H. Tang and L. Tassiulas was supported by NSF AI Institute Athena. The work of J. Zhang and J. Wang was supported in part by Tsinghua University--China Mobile Research Institute Joint Innovation Center. }
	}
	
	\maketitle
	
	\begin{abstract}
		In this work, we consider a status update system with a sensor and a receiver. The status update information is sampled by the sensor and then forwarded to the receiver through a channel with non-stationary delay distribution. The data freshness at the receiver is quantified by the Age-of-Information (AoI). The goal is to design an online sampling strategy that can minimize the average AoI when the non-stationary delay distribution is unknown. Assuming that channel delay distribution may change over time, to minimize the average AoI, we propose a joint stochastic approximation and non-parametric change point detection algorithm that can: (1) learn the optimum update threshold when the delay distribution remains static; (2) detect the change in transmission delay distribution quickly and then restart the learning process. Simulation results show that the proposed algorithm can quickly detect the delay changes, and the average AoI obtained by the proposed policy converges to the minimum AoI. 
	\end{abstract}
	
	
	\section{Introduction}
	The proliferation of real-time applications such as the remote surgery, virtual and augmented reality system has boosted the need for data freshness-oriented communication network design. To evaluate data freshness at the communication destination, the metric Age of Information (AoI) is introduced in \cite{yatesAoI}. When the transmission statistics of a channel is known, it is revealed that the AoI minimum transmission strategy will take a new sample and transmit it when data at the receiver is no longer fresh. 
	
	Learning the optimum sampling and scheduling strategy for minimizing the AoI performance in various communication scenarios are studied in \cite{kamlearning,vishrantlearning,tanglearning,yenerlearning,atay2021aging}. Assuming that the channel statistics remain unchanged but are unknown, \cite{kamlearning,tanglearning,yenerlearning} propose online learning methods to obtain the AoI minimum sampling and scheduling algorithms adaptively through stochastic bandits and reinforcement learning. Convergence results for a simple point-to-point communication link are provided in \cite{OnlineAoI2022thy,chih-chun}. Considering that channel conditions can be time-varying, \cite{atay2021aging} proposed an online learning algorithm that can achieve sub-linear regret under the worst-case channel state distribution. Considering that the data freshness performance is a function of the AoI, a learning algorithm to minimize the average AoI is proposed in \cite{vishrantlearning}. However, notice that the above research either assumes the channel to be static, or the channel conditions vary all the time, which are different from the piece-wise stationary channel conditions in real-world systems. 
	
	Transmission design in a changing environment is studied in \cite{guchange,tangchange}. By utilizing the angle domain sparse structure, a significance test of the subspace correlation is proposed in \cite{guchange} to detect channel changes in massive MIMO systems. To further maximize the transmission rate in a changing channel, a joint channel change and link rate selection algorithm is proposed in \cite{tangchange}. However, the threshold test in the above research requires the channel to be modeled by a specific distribution family, which is inaccurate when modeling the network delay data. 
	
	To overcome the algorithm design challenge when the channel delay distribution changes, we study the online freshness-oriented sampling in a point-to-point channel with semi-stationary delay distribution. The goal is to design a transmission strategy that can detect the delay distribution changes rapidly, and can converge to the optimum transmission policy under the current channel state quickly. We present an efficient delay change detection based on the Kolmogorov-Smirnov test from non-parametric statistics and integrate it with the online learning algorithm to minimize the average AoI. 
	
	The rest of the paper is organized as follows: Section II introduces the system model and optimization problem. Section III proposes a joint online learning and channel change point detection algorithm based on the Kolmogorov-Smirnov test. Simulation results are provided and analyzed in Section IV. Section V draws the conclusion. 
	\section{Problem Formulation}
	
	\subsection{System Model}\label{SysModel}

\begin{figure}[h]
		\centering
		\includegraphics[width=0.8\columnwidth]{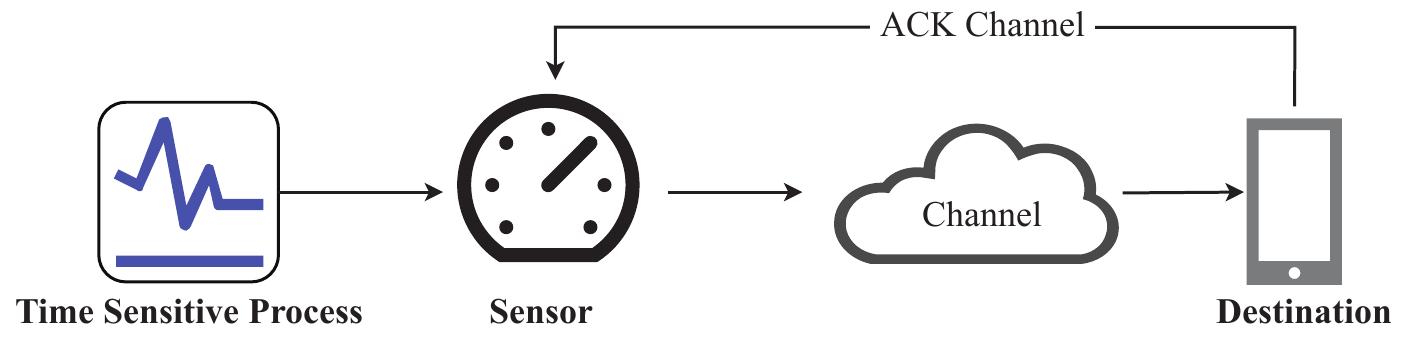}
		\caption{A point-to-point status update system.}
		\label{Fig:sysmodel}
	\end{figure}

	We consider a point-to-point status update system as depicted in Fig.~\ref{Fig:sysmodel}, where a sensor observes a time-sensitive physical process, samples status updates and meanwhile, sends them to the destination through a channel. Due to the limited channel capacity, only one packet can be transmitted at the same time.
	Once the transmission of an update finishes, an ACK signal will be sent to the sensor immediately. 
	
	Suppose the sensor can sample update packets at any time 
	$t \in \mathbb{R}^{+}$ and the sampling time-stamp of the $k$-th update packet is denoted by $S_{k}\in\mathbb{R}^+, k=1, 2, 3, \cdots$. The submitted packages will be served in a First-Come-First-Serve (FCFS) manner. The transmission delay in the communication channel is denoted by $D_k$. Therefore, the receiving time-stamp of the $k$-th update packet, denoted by $R_k$ can be computed by $R_k=\max\{R_{k-1}, S_k\}+D_k$. 
	
	We consider that the transmission channel is piece-wise stationary, i.e., the distribution of transmission delay $D_k$ keeps the same for a period of time, but can change after the stationary period. Specifically, we assume there exists $M$ channel state change points denoted by $\left\{ \tau_1, \tau_2, \cdots, \tau_M \right\}$ within the observation window $[0, T)$, and the delay distribution of $D_k$ within stationary period $[\tau_i, \tau_{i+1})$ follows distribution $\mathbb{P}_i$. To simplify the analysis, we assume the distribution of $D_k$ is determined at the time of $S_k$ and will not change\footnote{Such simplification is reasonable when $\tau_i - \tau_{i-1}$ is sufficiently large.}. We assume that the transmission delay is lower bounded by $D_{\mathsf{lb}}$.

	\subsection{Age of Information}
	We use AoI to evaluate the data freshness at the destination. By definition, AoI is the time elapsed since the freshest information stored at the destination is generated \cite{AoI2012Kaul}. Let
	$i(t):=\arg \max_{k\in\mathbb{N}}\left\{k\mid R_k \le t \right \}$
	be the index of the latest sample received by the destination before time $t$. The AoI at time $t$, denoted by $A(t)$ is:
	\begin{equation}
		A(t):=t-S_{i(t)}. 
	\end{equation}
	
\begin{figure}[h]
		\centering
		\includegraphics[width=1\columnwidth]{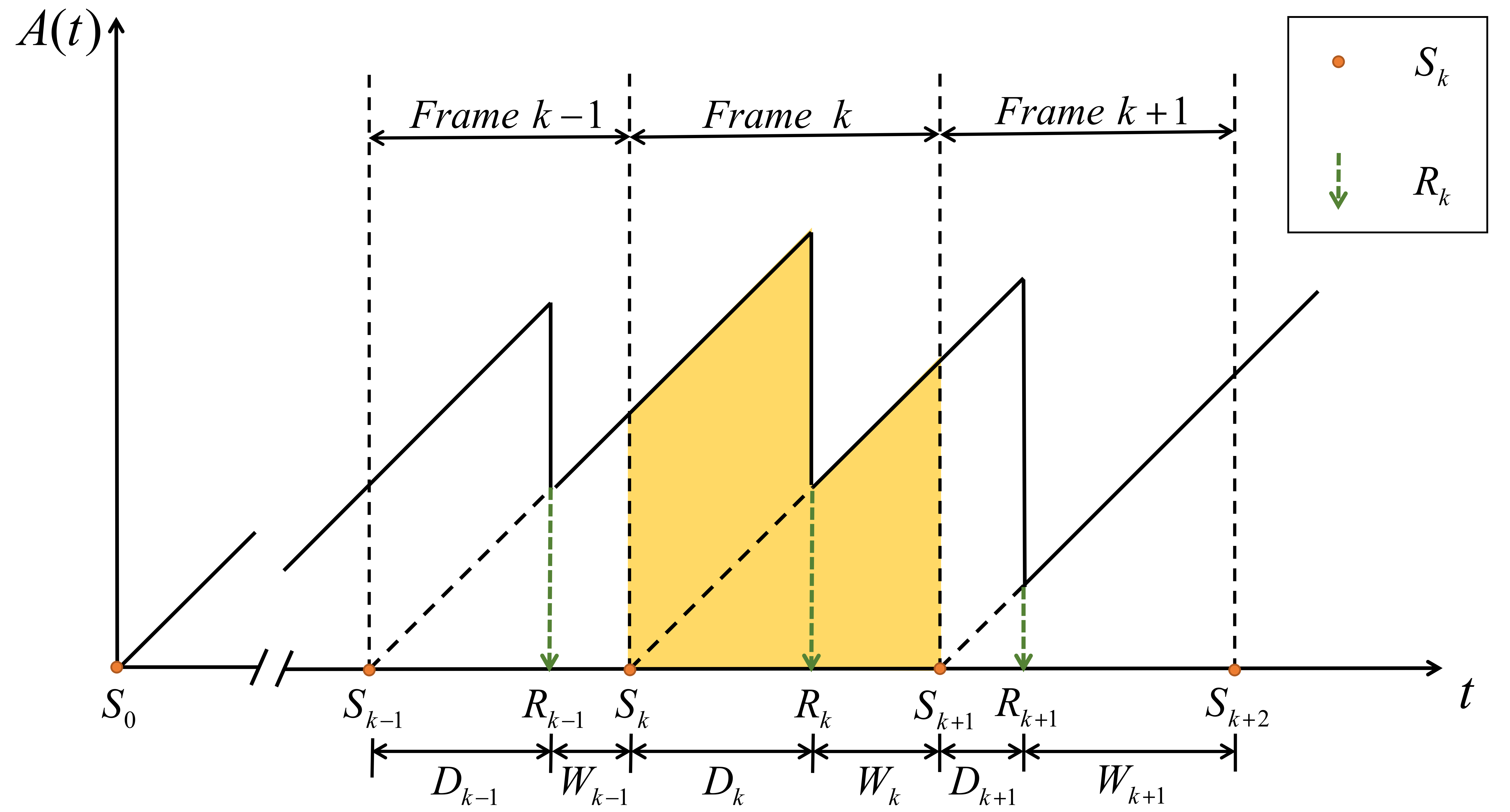}
		\caption{Illustration of the AoI evolution.}
		\label{Fig:AoI}
	\end{figure}

	A sample path of AoI evolution is depicted in Fig.~\ref{Fig:AoI}.

	\subsection{Optimization Problem}
	The goal of the sensor is to minimize the average AoI by designing a sampling strategy $\pi\triangleq\{S_1, S_2, \cdots\}$. Denote $\mathcal{H}_k:=\{(S_i, D_i)\}_{i=1}^{k}$ to be the set of historical sampling time-stamps and transmission delays.
	Our selection of the sampling time $S_k$ is based on historical sampling and transmission delays, i.e., $\mathcal{H}_{k-1}$. Future delay information $\{D_i\}_{i> k}$ is not accessible when determining the $k$-th sampling time. For a specific sampling policy $\pi$, the time-averaged AoI over the observation window $[0, T)$ can be computed as follows:
	
	\begin{equation}
		\overline{A}_{\pi}\triangleq \frac{1}{T}\mathbb{E}\left[\int_{0}^{T} A(t)\text{d}t\right]. \label{eq:avgAoIdef}
	\end{equation}
	
	\section{Problem Resolution}\label{sec:resolution}
	
	In this section, we first decompose the original problem into several sub-problems by assuming that the change points $\tau_i,\forall i$ are known in advance. Then we will reformulate each sub-problem into a renewal-reward process optimization \cite{OnlineAoI2022thy}. Finally, we propose a joint sampling and change detection strategy when the change points are unknown. 
	\subsection{Problem Reformulation and Decomposition}
	
	Since the interval of each change point remains unchanged, minimizing average AoI in time window $[0, T)$ is equivalent to minimizing average AoI within each interval $[\tau_m, \tau_{m+1})$, i.e., $
	\min_{\pi_m}\frac{1}{T}\int_{\tau_m}^{\tau_{m+1}} A(t)\text{d}t.$ When the interval $(\tau_{m+1}-\tau_m)$ is large, finding the optimum policy that minimizes the average AoI over $[\tau_m, \tau_{m+1})$ can be simplified by studying $\pi_m^\star$ that minimizes the average AoI over the infinite horizon:
	\begin{equation}a_{\pi_m^\star}:=\inf_{\pi}\limsup_{T\rightarrow\infty}\frac{1}{T}\mathbb{E}_{D\sim\mathbb{P}_m}\left[\int_{\tau_m}^{\tau_{m+1}} A(t)\text{d}t\right].\label{eq:pb2}
	\end{equation}
	
	We then focus on one such problem and neglect the subscript $m$ henceforth. Previous work \cite{Update2017Sun} has illustrated that the optimum policy that achieves the minimum AoI is a stationary policy that selects a waiting time $W_k$ after receiving the $k$-th ACK. We then limit our search of $\pi^\star$ within such a set of waiting policies denoted by $\Pi_W$. 
	
	To facilitate the average AoI computation of a stationary sampling policy, denote the $k$-th frame to be the time interval between $S_{k}$ and $S_{k + 1}$. The cumulative AoI in frame $k$ is denoted by $X_k:=\int_{S_k}^{S_{k+1}}A(t)\text{d}t$. The calculation of $X_k$ can be converted into the computing the area sum of the colored parallelogram and triangle in Fig.~\ref{Fig:AoI}, i.e.,
	
	\begin{equation*}
		X_k=(D_{k-1}+W_{k-1})D_k+\frac{1}{2}(D_k+W_k)^2.
	\end{equation*}
	
	Then the cumulative AoI over interval $[0, S_{K+1})$ can be rewritten as a sum of $X_k$, i.e.,
	\begin{align}
		&\mathbb{E}\left[\int_{0}^{S_{K+1}} A(t)\text{d}t\right]=\mathbb{E}\left[\sum_{k=1}^KX_k\right].\label{eq:aoidecomp}
	\end{align}

		Previous work \cite{Update2017Sun} and \cite{Tsai2020AoIRT} have already proven that it is sufficient to consider the stationary deterministic policy whose waiting time is a stationary mapping from transmission delay, i.e., $W_k = w(D_k)$ where function $w:[0,\infty)\to[0,\infty)$. Let $L_2$ denote the Lebesgue space. Then searching for the optimum policy $\pi^\star$ to \eqref{eq:pb2} can be reformulated into the following problem:
		\begin{pb}[Renewal-Reward Process Reformulation]
			\label{pb:r-rpr}
			\begin{align}
				\inf_{w\in L_2}\frac{ \mathbb{E}_{\mathbb{P}}\left[\frac{1}{2} (D+w(D))^2 \right]} {\mathbb{E}_{\mathbb{P}}\left[ D+w(D)) \right]}+\overline{D}. 
			\end{align}
		\end{pb}

		\subsection{Optimal Sampling for Stationary Sub-Problem}
		The optimum policy to the renewal-reward process $\pi^\star$ selects waiting time $w(\cdot)$ as follows:
		\begin{equation}
			w(d)=(\gamma^\star-d)^+,
		\end{equation}
		where $\gamma^\star=a_{\pi^\star}-\overline{D}$, $a_{\pi^\star}$ is the average AoI obtained by the optimum policy. To compute the optimum parameter $\gamma^\star$ that resolves Problem~\ref{pb:r-rpr}, we maintain a sequence $\{\gamma_k\}$ that samples and update our guessing about $\gamma^\star$ in each frame $k$ as follows:
		\begin{itemize}
			\item \textbf{Initialization:} $\gamma_1 = 0$. 
			
			\item \textbf{Sampling:} After receiving the ACK signal of the $k$-th update packet, the waiting time $W_k$ is chosen based on the current estimation $\gamma_k$:
			\begin{equation*}
				W_k = \left(\gamma_k- D_k \right)^+. 
			\end{equation*}
			
			\item \textbf{Update:} $\gamma_k$ is updated through Robbins-Monro algorithm \cite{robbins1951stochastic}:
			\begin{equation*}
				\gamma_{k+1} = \left(\gamma_k + \eta_k (Q_k-\gamma_k L_k)\right)^+,
			\end{equation*}
			where $Q_k=\frac{1}{2}(D_k + W_k)^2$, $L_k = D_k + W_k$, and $\eta_k$ is the step size:
			\begin{align}
				\eta_k = \begin{cases}
					\frac{1}{2D_{\mathsf{lb}}}, & k=1; \\
					\frac{1}{(k+2)D_{\mathsf{lb}}}, & k \ge 2.
				\end{cases}\label{eq:eta_k}
			\end{align}
			
		\end{itemize}

		\subsection{Change Point Detection}
		The optimal $\gamma_m^\star$ within each interval $[\tau_m, \tau_{m+1})$ depends on the delay statistics $\mathbb{P}_m$ and therefore, may be different from $\gamma_{m'}$ in another interval $m'$. The decreasing step sizes selected in \eqref{eq:eta_k} implies,  as time moves on, the learning rate is becoming smaller. This motivates us to detect these channel change points to speed up the convergence of the algorithm once the channel delay distribution changes.
		
		In this work, since we do not restrict the delay distribution $\mathbb{P}_m$ to be a specific parameter family, we apply the Kolmogorov-Smirnov (KS) test \cite{smirnov1948table} from non-parametric detection to determine whether the current delay distribution has changed. Due to the limited amount of available data, to ensure test effectiveness, the threshold of determining whether a channel change point happens is set through bootstrapping. After observing the delay $D_k$ in each frame $k$, we perform the change point detection test by comparing the empirical distribution of the last $n$ samples and the second last $n$ samples using the KS test as shown in Algorithm \ref{alg:KS}.
		
		\begin{algorithm}[h]
			\caption{Two-sample Kolmogorov–Smirnov test} \label{alg:KS}
			\begin{algorithmic}[1]
				\STATE \textbf{Input}: Confidence level $\alpha\in(0, 1)$
				
				\STATE \textbf{Dataset Construction}:
				\begin{subequations}\begin{align}\label{eqD1}
						\mathcal{D}_1 = \{D_{k}, D_{k-1}, ..., D_{k-n+1}\},\\
						\label{eqD2}
						\mathcal{D}_2 = \{D_{k-n}, D_{k-n-1}, ..., D_{k-2n+1}\}.
					\end{align}
				\end{subequations}

				\STATE Divide $[0, D_{max}]$ into intervals $[x_i, x_{i+1})$, where we use $\mathcal{X}:=\{x_1, \cdots, x_N\}$ to be the set of cutting points. 
				
				\begin{subequations}\begin{align}\label{eq:F1n}
						F_{1}(x)=\frac{1}{n} \sum_{d\in\mathcal{D}_1} \mathbbm{1}(d\in\left( -\infty,x \right]), \forall x\in\mathcal{X},\\
						\label{eq:F2n}
						F_{2}(x)=\frac{1}{n} \sum_{d\in\mathcal{D}_2} \mathbbm{1}(d\in\left( -\infty,x \right]), \forall x\in\mathcal{X}.
					\end{align}
				\end{subequations}
				
				\STATE Compute
				\begin{equation}
					\Delta=\max_{x\in\mathcal{X}}{|F_{1}(x)-F_{2}(x)|}. 
				\end{equation}
				
				\FOR{$i=1,2,\cdots,R$}
				\STATE Bootstrap $\hat{\mathcal{D}}_{i, 1}$ and $\hat{\mathcal{D}}_{i, 2}$ of $n$ samples from dataset $\mathcal{D}_1$ and $\mathcal{D}_2$. 
				\STATE Compute the empirical distribution $\hat{F}_{1, i}$, $\hat{F}_{2, i}$ similar to \eqref{eq:F1n} and \eqref{eq:F2n}.
				
				\STATE Compute $\Delta_i=\max_{x\in\mathcal{X}}\left|\hat{F}_{1, i}(x)-\hat{F}_{2, i}(x)\right|$. 
				\ENDFOR
				
				\STATE Assign $\delta$ the $\lfloor \alpha R\rfloor$ largest value in sequence $\{\Delta_i\}$.
				\IF{$\Delta>\delta$}
				
				\STATE Output: Change detected.
				\ELSE
				\STATE Output: No change happens. 
				
				\ENDIF
				
			\end{algorithmic}
		\end{algorithm}
		

		\subsection{Algorithm Integration}
		Now we are ready to propose our algorithm by combining the aforementioned online sampling and change point methods together. First we initialize the estimation $\gamma_1$ and use $\tau$ to record the latest detected change point. Then after receiving the ACK of the $k$-th packet, we perform the joint online sampling and change point detection algorithm (algorithm \ref{alg:online+CD}):
		\begin{itemize}
			\item \textbf{Online Sampling:} Choose the waiting time according to Eq.~\eqref{eq:W_k}, and update the estimation $\gamma_k$ along with $\eta_k$ in step 13.
			
			\item \textbf{Change Point Detection:} If there are more than $2m$ packets having been transmitted since the latest change point, then run the change point detection algorithm. If a new change is claimed, then update the latest change and reset the online sampling algorithm, i.e., reset $\gamma_k$ and $\eta_k$.
		\end{itemize} 
		\begin{algorithm}
			\caption{Joint online sampling and change point detection} \label{alg:online+CD}
			\begin{algorithmic}[1]
				\STATE \textbf{Initialization}: $\gamma_1=0,~\tau=0$.
				\FOR{$k=1,2,\cdots,K$}
				\STATE Receive the ACK signal of the $k$-th update packet, then compute the transmission delay $D_k$.
				\STATE Choose the waiting time $W_k$ using current $\gamma_k$:
				\begin{equation}\label{eq:W_k}
					W_k = (\gamma_k -D_k)^+.
				\end{equation}
				\STATE Compute $Q_k = \frac{1}{2}(D_k+W_k)^2$ and $L_k = D_k+W_k$.
				\IF{$k-\tau > 2m$}
				\STATE Run the change point algorithm Alg.~\ref{alg:KS}.
				\IF{claim a change}
				\STATE Record the change point $\tau = k$.
				\STATE Reset the algorithm $\gamma_k = 0$.
				\ENDIF
				\ENDIF
				\STATE Update $\gamma_{k+1} = [\gamma_k + \eta_{k-\tau} (Q_k-\gamma_k L_k)]^+$, where $\eta_k$ is chosen according to Eq.~\eqref{eq:eta_k}.
				\ENDFOR
			\end{algorithmic}
		\end{algorithm}


		\section{Simulation Results}
		In this section, we validate the performance gain of the proposed joint learning and channel change detection algorithm via numerical simulations. We compare the expected time-averaged AoI of the proposed $\pi_{\text{online-KS}}$ with: (1) The zero-wait sampling policy $\pi_{\text{zw}}$ that selects $w(\cdot)\equiv 0$; (2) The AoI minimum sampling policy $\pi^\star$ when the delay distribution is known; (3) Online AoI minimization policy $\pi_{\text{online}}$ \cite{tanglearning} without a change point detection algorithm. 
		
		In simulations, we consider an observation window with $T=3\times 10^5$ time slots. Consider that there are $M=2$ change points with $\tau_1=1\times 10^5$ and $\tau_2=2\times 10^5$. For simplicity, consider that the transmission delay follows the lognormal distribution parameterized by $\mu$ and $\sigma$, i.e., the probability density function 
		\[p(d)=\frac{1}{\sigma\sqrt{2\pi}}\exp\left(-\frac{(\ln d-\mu)^2}{2\sigma^2}\right).\]
		Assume that the parameter set \{$\mu$, $\sigma$\} of delay distributions within $[0, \tau_1)$, $[\tau_1, \tau_2)$, $[\tau_2, T)$ are \{$0.3$, $1.25$\}, \{$-1.0$, $1.00$\} and \{$-0.2$, $1.10$\}, respectively. Due to computational complexity, the threshold of determining whether a channel change point happens is set to be a fixed value.

		
		
		
		
\begin{figure}[H]
			\centering
			\includegraphics[width=1\columnwidth]{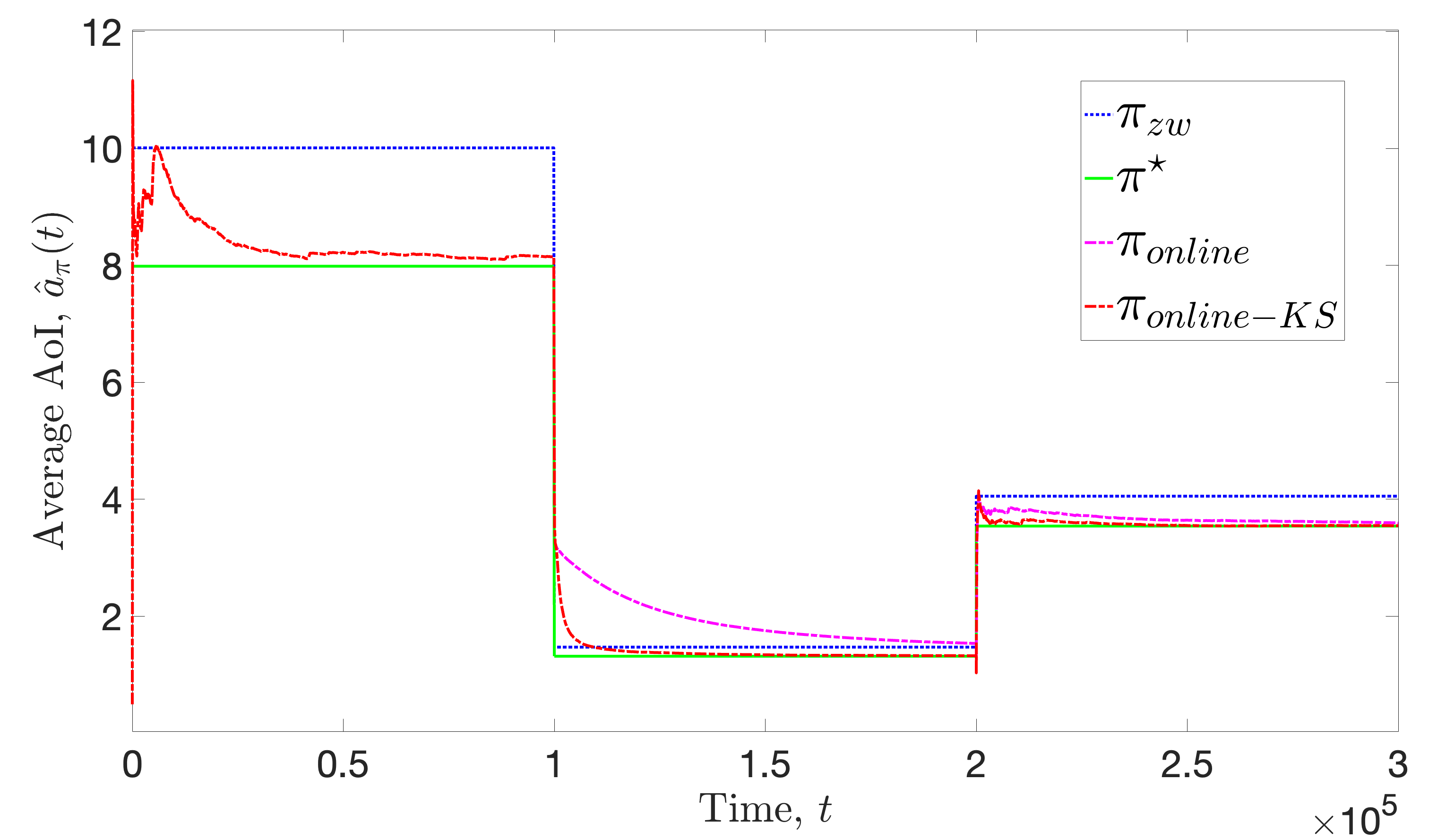}
			\caption{Average AoI Comparisons of Various Algorithms}
			\label{Fig:Consequence}
		\end{figure}

		To validate the performance gain of the proposed algorithm, let $\hat{a}_{\pi}(t)$ be the expected average AoI using policy $\pi$ from $\tau_i$, the latest channel change point at time $t$, i.e., 
		\[\hat{a}_{\pi}(t)=\frac{1}{t-\tau_i}\int_{\tau_i}^tA(\tau)\text{d}\tau. \]
		We plotted $\hat{a}_{\pi}(t)$ for different policies by taking the average of 30 runs. The simulation consequence is depicted in Fig.\ref{Fig:Consequence}. 
		
		According to Fig.~\ref{Fig:Consequence}, the average AoI of the proposed algorithm converges to the average AoI obtained by the optimum offline policy $\pi^\star$ in each stationary period. Moreover, the average AoI obtained by $\pi_{\text{online-KS}}$ is smaller than policy $\pi_{\text{online}}$, the online learning without a change point detection algorithm. It is because when the channel delay distribution changes, our proposed algorithm is able to detect the change rapidly. Therefore, it can immediately abandon the historical data, 'restarting' itself quickly to better fit into the new channel. Compared with policy $\pi_{\text{online}}$ without a change point detection algorithm, the proposed $\pi_{\text{online-KS}}$ has a faster convergence speed. 
		
		\section{Conclusions}
		
		In this paper, we study age-optimal sampling in piece-wise stationary environment. We propose a joint online sampling and change-point detection algorithm by decoupling the initial problem into several sub-problems during each stationary period.  We then observe the performance of our proposed algorithm through simulation, whose results show that our strategy achieves a lower average AoI.

		\bibliographystyle{IEEEtran}
		\bibliography{bibfile}
	\end{document}